\documentclass[amsmath,superscriptaddress,showpacs,prb,twocolumn] {revtex4-1}
\usepackage{bm}
\usepackage{enumerate}
\usepackage{graphicx}
\usepackage[dvips]{epsfig}
\usepackage{epsf}
\usepackage{amsfonts}

\begin{document}
\title{Spin Hall and Nernst effects of Weyl magnons}

\author{Vladimir A. Zyuzin} 
\affiliation{Department of Physics and Astronomy, Texas A$\mathrm{\&}$M University, College Station, Texas 77843-4242, USA}
\affiliation{Department of Physics and Astronomy and Nebraska Center for Materials
and Nanoscience, University of Nebraska, Lincoln, Nebraska 68588,
USA}
\author{Alexey A. Kovalev}
\affiliation{Department of Physics and Astronomy and Nebraska Center for Materials
and Nanoscience, University of Nebraska, Lincoln, Nebraska 68588,
USA}
\begin{abstract}
In this paper, we present a simple model of a three-dimensional insulating magnetic structure which represents a magnonic analog of the layered electronic system described in [Phys. Rev. Lett. {\bf 107}, 127205 (2011)]. In particular, our model realizes Weyl magnons as well as surface states with a Dirac spectrum.
In this model, the Dzyaloshinskii-Moriya interaction is responsible for the separation of opposite Weyl points in momentum space.
We calculate the intrinsic (due to the Berry curvature) transport properties of Weyl and so-called anomalous Hall effect (AHE) magnons.
The results are compared with fermionic analogs. 
\end{abstract}

\maketitle

\section{Introduction}
Recently, studies of intrinsic (topological) properties of fermionic systems have received tremendous interest from the research community. 
Some of these studies have concentrated on the quantum Hall effect,\cite{LaughlinPhysRevB.23.5632,TKNN_PhysRevLett.49.405} Chern insulators,\cite{HaldanePhysRevLett.61.2015} topological insulators,\cite{VolkovPankratovJETPlett1985, KaneMelePhysRevLett.95.146802,HasanKaneRevModPhys.82.3045,Bernevig.Hughes.ea:S2006} and Dirac (Weyl)\cite{Murakami2007NJP,WanTurnerVishwanathSavrasovPhysRevB.83.205101,BurkovBalentsPhysRevLett.107.127205} semimetals. 
Of particular interest are the transport properties and transitions between various topological phases. 
The Berry curvature\cite{Berry1984} plays an important role in descriptions of the intrinsic transport properties such as the Hall, Nernst, and axial or chiral current responses. \cite{RevModPhys.82.1959}

In a phase transition that separates two insulating phases with different topological numbers, a semimetal phase necessarily occurs. 
This semimetal phase is characterized by a band touching and, in general is called the Dirac semimetal, a condensed-matter analog of relativistic fermions. 
Under breaking of either time-reversal or inversion symmetry, opposite chiralities separate either in momentum or energy, and in this way the so-called Weyl semimetal is stabilized. 
This scenario is realized in an analytical model presented in Ref.~[\onlinecite{BurkovBalentsPhysRevLett.107.127205}].
The Weyl semimetal phase is of interest as it exhibits the anomalous Hall effect (AHE), surface Fermi arcs, and chiral anomaly driven responses.

Similar topological effects are recognized in magnetic insulating systems. 
Due to a combination of the underlying lattice geometry and Dzyaloshinskii-Moriya interaction (DMI),\cite{Dzyaloshinsky:JoPaCoS1958, Moriya:PR1960} the magnon bands can acquire a non-trivial Berry curvature and non-vanishing Chern numbers. \cite{Matsumoto.Murakami:PRL2011, Shindou.Matsumoto.ea:PRB2013,Zhang.Ren.ea:PRB2013,Shindou.Ohe.ea:PRB2013,Mook.Henk.ea:PRB2014,Mook.Henk.ea:PRB2014a} 
As in the case of fermions, magnons can exhibit spin Nernst\cite{KovalevZyuzin2016,RanOkamotoXiao2017,ZyuzinKovalev2016} and thermal Hall responses,\cite{Katsura.Nagaosa.ea:PRL2010,Matsumoto.Murakami:PRL2011,
Zhang.Ren.ea:PRB2013,Matsumoto.Shindou.ea:PRB2014,Lee.Han.ea:PRB2015,Owerre2016JAP} and induce dissipative torques\cite{KovalevZyuzin2016} on the magnetic order. 
For example, the thermal Hall effect carried by magnons has been experimentally observed in insulating collinear ferromagnets with pyrochlore crystall structure.\cite{Onose.Ideue.ea:S2010,Ideue.Onose.ea:PRB2012} 
The spin Nernst effect carried by magnons\cite{RanOkamotoXiao2017,ZyuzinKovalev2016} was recently observed\cite{ShiomiTakashimaSaitoh2017} in an antiferromagnet. 

Different magnetic models have been proposed for realizations of the aforementioned intrinsic effects. 
These include two-dimensional kagome\cite{MishchenkoStarykhPhysRevB.90.035114, Mook.Henk.ea:PRB2014a, KovalevZyuzin2016} and honeycomb\cite{OwerreJPCM2016,Owerre2016JAP,PhysRevLett.117.227201,Wang.Su.ea:PRB2017} magnets as well as pyrochlore\cite{Ideue.Onose.ea:PRB2012,PhysRevLett.118.177201} and layered structures. 
By tuning exchange parameters some of the above magnetic systems reveal magnons described by a Weyl spectrum.\cite{WeylMagnons2016NatComm, PhysRevLett.117.157204, PhysRevB.95.224403,0256-307X-34-7-077501,Su.Wang:PRB2017,2017arXiv170802948J,OwerreArxiv2017a,OwerreArxiv2017b}  

\begin{figure} \centerline{\includegraphics[clip, width=0.9  \columnwidth]{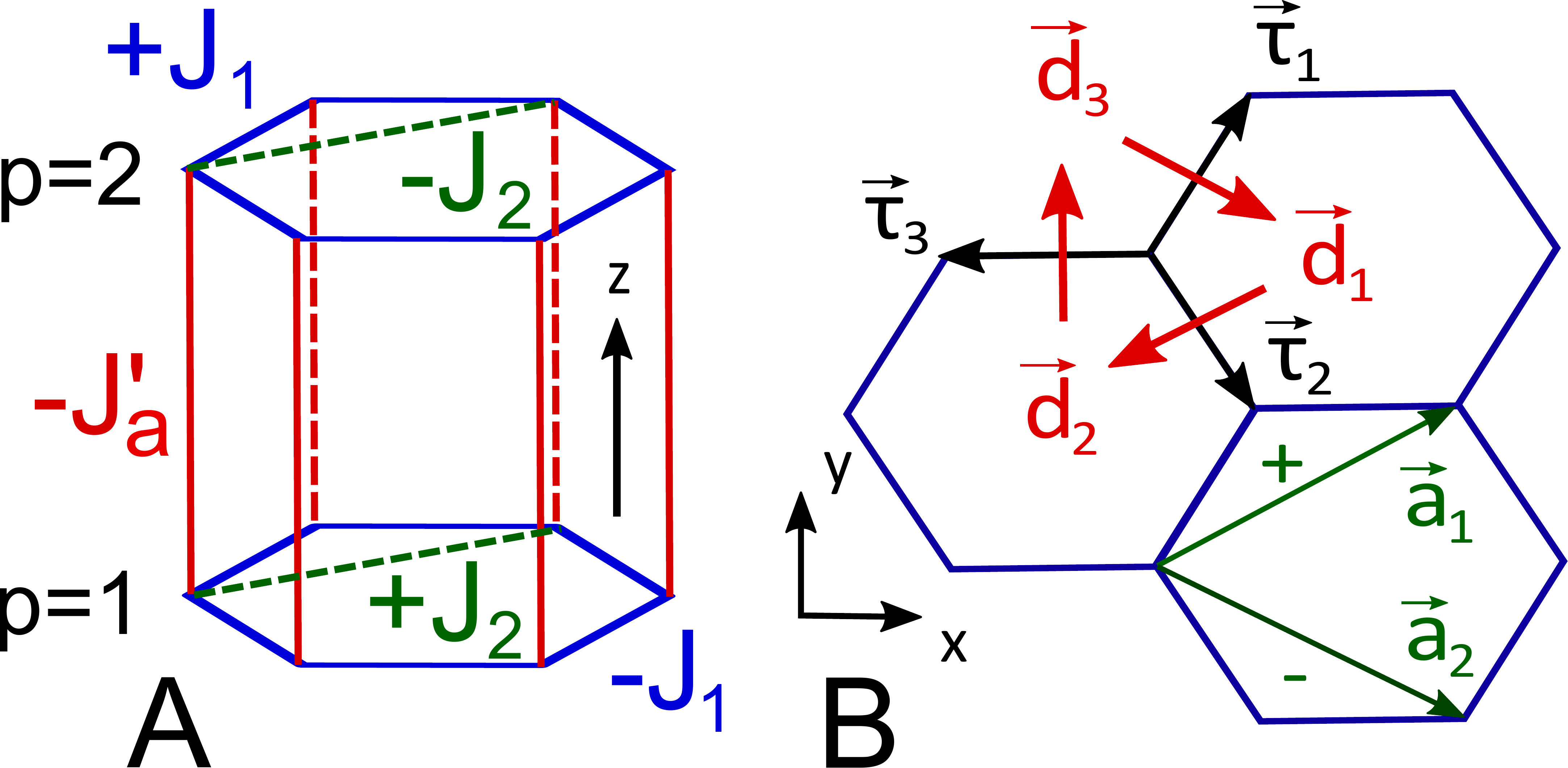}}

\protect\caption{(Color online) A. Unit cell of the system. Constants $J_{1}$, $J_{2}$, and $J_{\mathrm{a}}^{\prime}$ denoting the exchange interactions are all chosen to be positive.
B. Schematics of the honeycomb lattice parameters used in the derivation of the non-interacting magnon spectrum. 
Vectors connecting nearest neighbors are  ${\bm \tau}_{1} = \frac{1}{2}(\frac{1}{\sqrt{3}},1 )$,
 ${\bm \tau}_{2} = \frac{1}{2}(\frac{1}{\sqrt{3}},-1)$, and ${\bm \tau}_{3} = \frac{1}{\sqrt{3}}(-1,0 )$. 
Vectors ${\bf a}_{1} = \frac{1}{2}(\sqrt{3},1)$, and ${\bf a}_{2} = \frac{1}{2}(\sqrt{3}, -1)$ are used in deriving second-nearest neighbor exchange interaction and DMI. Green $\pm$ signs denote the signs of the ${\bf D}_{(ij)}^{[\mathrm{z}]}$ vector for the $(ij)$ link defined by green arrows.}

\label{fig1}  

\end{figure}

In this paper, we propose a new model that realizes Weyl magnons and the magnon analog of Fermi arcs.      
The model contains interchanging layers of honeycomb ferromagnets and antiferromagnets (see Fig. \ref{fig1}). 
This is needed to establish the opposite chiralities of magnons.  
We show that in this model the magnon spectrum and topology qualitatively resemble those considered in Ref.~[\onlinecite{BurkovBalentsPhysRevLett.107.127205}] and [\onlinecite{PhysRevB.85.165110}] for fermions. 
For example, by varying inter-layer exchange parameters the nodal-line spectrum of magnons emerges (see Fig. \ref{fig}). 
Furthermore, we observe a magnon surface states with the Dirac spectrum. 
These surface states might hybridize with the bulk states as they are shifted in energy. 
Our model can also be interpreted as a magnon analog of the 3D Shockley model.\cite{PhysRevB.86.075304,Pershoguba.Banerjee.ea:PRX2018} 
When the DMI is switched on, either Weyl magnons or the magnon analog of stacked two-dimensional anomalous Hall effect layers (the so-called AHE magnons) are obtained.

We study intrinsic (due to the Berry curvature) spin transport properties of the Weyl and AHE magnons. 
Magnon pumping due to magnetization dynamics was discussed in Ref.~[\onlinecite{KovalevZyuzinLiPhysRevB.95.165106}]. 
Importantly for the present paper, one can draw an analogy between fermions responding to electric field and magnons responding to magnetization dynamics.  
As mentioned in Ref.~[\onlinecite{BurkovBalentsPhysRevLett.107.127205}], in the case of a Weyl semimetal the AHE is semi-quantized, and it is proportional to the splitting of Weyl points in momentum space.
In this paper, we show that the vicinity of the Weyl points leads to the magnon-driven spin current  proportional to the splitting of Weyl points. However, other regions of the Brillouin zone also contribute to the response.
For the AHE magnons, at small temperatures we recover the results of Ref.~[\onlinecite{KovalevZyuzinLiPhysRevB.95.165106}]; that is, the response is proportional to the DMI strength, and it is a response of a number of stacked layers of two-dimensional Chern magnons.  
At higher temperatures, the response from the Weyl points acquires an extra logarithmic factor corresponding to the energy cutoff. 
In both cases the responses are temperature dependent, vanishing at zero temperature.

This paper is organized as follows. In section \ref{section1}, we briefly review the Shockley model given in Ref.~[\onlinecite{PhysRevB.86.075304}]. 
Then, we construct a more general model that contains the Weyl and AHE magnons. We discuss how different phases emerge as the parameters of the model, such as DMI and exchange interactions, change.
In section \ref{section2}, we study the intrinsic responses of the Weyl and AHE magnons, both analytically and numerically. 
For the analytical calculations, we adopt a simplified model that captures the contribution from the Weyl points. 
In the Appendix, we give details of the calculations.

\section{Topological magnons in layered systems} 
\label{section1}
\subsection{3D Shockley-like model}
Before we formulate our model of Weyl magnons, we give here a brief description of the 3D Shockley-like model introduced in Refs.~[\onlinecite{PhysRevB.86.075304,Pershoguba.Banerjee.ea:PRX2018}]. 
In such Shockley-like models one can obtain different topological phases with surface states. 
It is also known that a Weyl semimetal occurs at the phase transitions between the topological phases.\cite{Murakami2007NJP} 
Therefore, in the magnon version of the Shockley-like model one can expect magnon analogs of known topological phases including the   Weyl phase, which is of particular interest to us. 
The Shockley-like model is described by the Hamiltonian
\begin{align}
\label{eq:Shockley}
H &=  \left(
\begin{array}{cc}
 h({\bf k_{\parallel}}) & t(k_z,{\bf k_{\parallel}}) \\
t^{*}(k_z,{\bf k_{\parallel}}) & -h({\bf k_{\parallel}}) \\
\end{array}
\right),
\end{align}
where two types of interchanging layers are described by $\pm h({\bf k_{\parallel}})$ and the interlayer hopping amplitudes are described by $t(k_z,{\bf k_{\parallel}})$.
 Note that $h({\bf k_{\parallel}})$ could in principle correspond to a matrix, e.g., due to the spin or sublattice degrees of freedom. 
Taking $t(k_z,{\bf k_{\parallel}})=t_1({\bf k_{\parallel}})e^{-i k_z}+t_2({\bf k_{\parallel}}) e^{i k_z}$, one can obtain that such a model can describe surface states when $|t_1({\bf k_{\parallel}})|<|t_2({\bf k_{\parallel}})|$, where the layers have to be interrupted at the $t_2({\bf k_{\parallel}})$ bond. 
In $(k_x,k_y)$ regions where such a condition is satisfied the surface states are described by the spectrum $h({\bf k_{\parallel}})$, and can contain a Dirac cone. 
As we will show below, our model of Weyl magnons given by Eq.~(\ref{eq:matrix}) without the DMI corresponds to the model in Eq.~(\ref{eq:Shockley}).

\begin{figure} \centerline{\includegraphics[clip, width=0.9  \columnwidth]{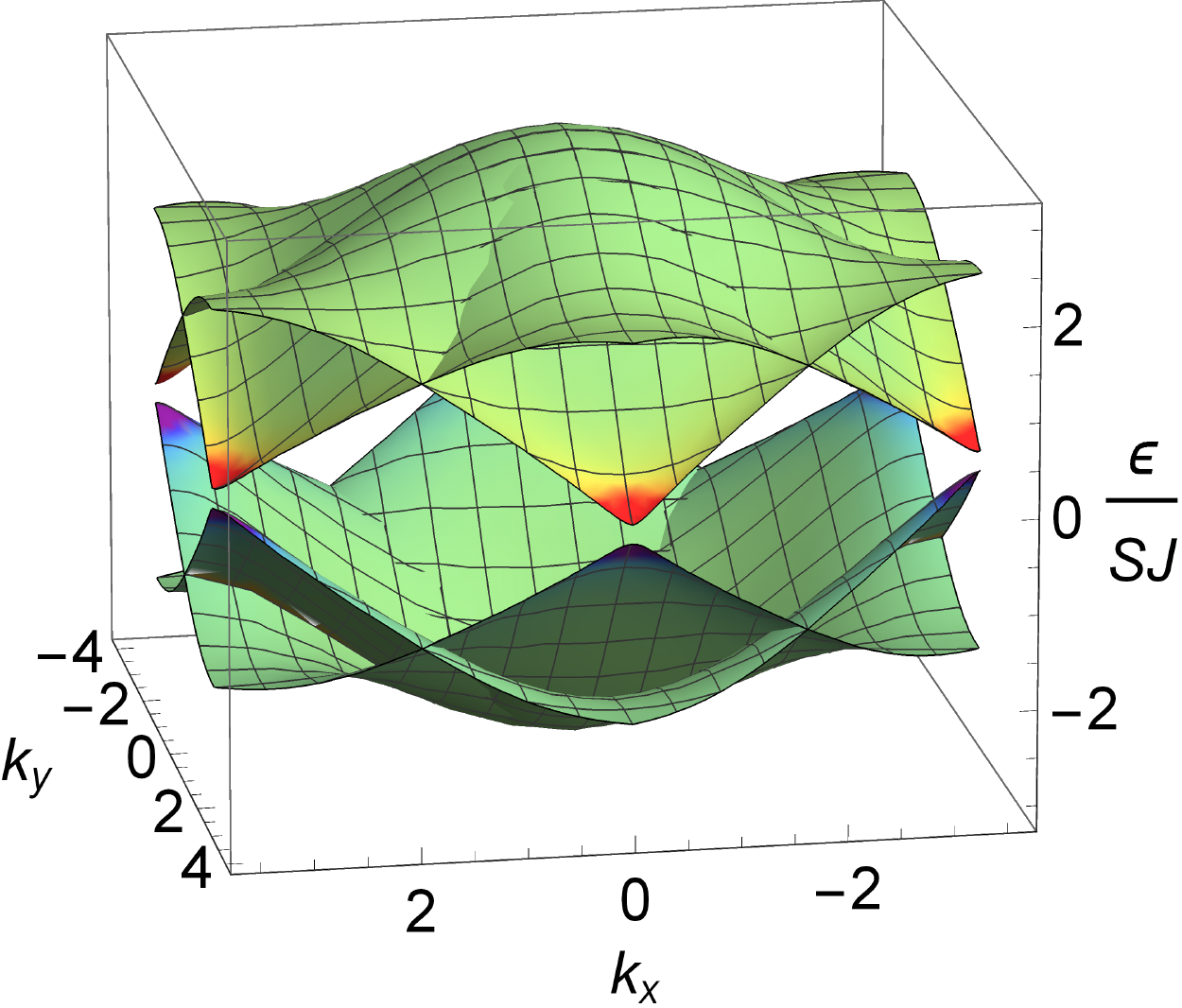}}

\protect\caption{(Color online) The spectrum of Weyl magnons for the case $\lambda_{\boldsymbol k}=0$ described in Eq. (\ref{Berry}) at $k_z=\mathrm{const}$ corresponding to the vicinity of the Weyl points. 
The Weyl points are located at the ${\bf K}$ and ${\bf K}^\prime$ points at $k_{z} = \pm \arccos\left( \frac{1}{2\sqrt{ t_{\mathrm{a}}t_{\mathrm{e}} }} \sqrt{3\Delta^2 - 1 } \right)$. The red and blue colors correspond to positive and negative Berry curvatures, respectively.}

\label{fig}  

\end{figure}

\subsection{A realization with magnons}
We study spins located at the sites of a three dimensional lattice made of honeycomb layers stacked in the AA type. 
The unit cell of the system contains two honeycomb layers (see Fig.~\ref{fig1}A). 
For the bottom layer of the unit cell the first and second nearest neighbor exchange couplings are assumed to be ferromagnetic and antiferromagnetic, respectively. 
For the top layer the types of exchange couplings are switched (see Fig.~\ref{fig1}A). 
For simplicity, the exchange coupling between the layers, $J^{\prime}_{\mathrm{a}}$, within the unit cell is chosen to be ferromagnetic.
We also include the Dzyaloshinskii-Moriya interaction (DMI) and an external magnetic field. 
A three dimensional system is obtained by translating the unit cell in the $z-$direction with the inter unit cell exchange coupling $J^{\prime}_{\mathrm{e}}$. 
The spin Hamiltonian is 
\begin{align}\label{ExchangeHamiltonian}
H &= \sum_{m p \langle ij \rangle } (-1)^{p} J_{1} {\bf S}_{mpi}{\bf S}_{mpj} - (-1)^{p} J_{2} {\bf S}_{mpi}{\bf S}_{mpj} 
\\
&
+ \sum_{m p \langle\langle ij \rangle\rangle } {\bf D}_{(ij)}^{[\mathrm{z}]} \left[{\bf S}_{mpi} \times {\bf S}_{mpj} \right]
- \mu_{\mathrm{B}}  \sum_{mki} {\bf B}{\bf S}_{mpi}
\nonumber
\\
&
+ \sum_{m p \langle ij \rangle } (-1)^{p}{\bf D}_{(ij)}^{[\mathrm{R}]} \left[ {\bf S}_{mpi} \times {\bf S}_{mpj} \right]
\nonumber
\\
&
- \sum_{m  i} J^{\prime}_{\mathrm{a}}    {\bf S}_{m 1 i}{\bf S}_{m 2 i} 
- \sum_{m  i} J^{\prime}_{\mathrm{e}}    {\bf S}_{m 2 i}{\bf S}_{m+1, 1 i} \,,
\nonumber
\end{align}
where the constants $J_{1}$, $J_{2}$, $J_{\mathrm{a}}^{\prime}$, and $J^{\prime}_{\mathrm{e}}$ denoting the exchange coupling are chosen to be positive 
and index $p=1,2$ denotes the bottom and top layers of the unit cell. 
The unit cell is translated in the $z-$direction, and $m = 1,2,3,\dots$ denotes the number of the unit cell. 
The vector ${\bf D}_{(ij)}^{[\mathrm{z}]} = D^{[\mathrm{z}]}{\bf e}_{z}\nu_{ij}$ is the out of plane second-nearest neighbor DMI with $\nu_{ij}=\pm$ signs for an $(ij)$ link shown in green in Fig.~\ref{fig1}B and 
the vector ${\bf D}_{(ij)}^{[\mathrm{R}]} = D^{[\mathrm{R}]} {\bf d}_{l}$ is the in-plane DMI of the Rashba type. 
Index $l=1,2,3$ denotes an $(ij)$ link on a lattice corresponding to a ${\bm \tau}_{l}$ vector (see Fig. \ref{fig1}B) along which the spins interact.
For a link $l$ the Rashba DMI vectors are ${\bf d}_{1} = \frac{1}{2}(\sqrt{3},-1)$, ${\bf d}_{2} = \frac{1}{2}(-\sqrt{3},-1)$, and ${\bf d}_{3} = (0,1)$, defined in red in Fig.~\ref{fig1}B. 
Note that the sign of the Rashba DMI for the same bond is opposite in the bottom and top layers. 
The reason for such a choice of DMI will be seen in the next section, where we study the responses of the system.
Such DMI can be achieved by placing an extra non-magnetic charged layer between the top and bottom layers, such that the Rashba spin-orbit coupling is generated with opposite signs in the top and bottom layers.   
Note that there might also be a second-nearest neighbor in-plane DMI of the Rashba type. 
We omitted it as it does not lead to qualitatively different physical picture and is smaller than the first-neighbor Rashba DMI. 
All DMIs are small, such that $\frac{D^{[\mathrm{z}]}}{J_{1}} \ll 1$ and $\frac{D^{[\mathrm{R}]}}{J_{1}} \ll 1$. 
We assume that the magnetic field ${\bf B}$ is above the saturation value so that all spins align with the field.
The direction of the field and hence of the magnetization is assumed to be general, namely $(m_{x},m_{y},m_{z})$, however, with the main component in the $z-$ direction, i.e. $m_{z} \gg m_{x},m_{y}$.
Alternatively, the magnetic anisotropy in the $z-$ direction could also be used instead of the magnetic field to align spins in the $z-$ direction.

We are now ready to study the magnons, fluctuations around the magnetization direction. 
The unit cell contains four elements, and a set of four boson operators is needed to describe the magnons.
We perform the Holstein-Primakoff transformation, $S_{n}^{z}({\bf r}) = S - a_{n}^{\dag}({\bf r})a_{n}({\bf r})$ and $S^{+}_{n}({\bf r}) = S_{n}^{x}({\bf r}) + iS_{n}^{y}({\bf r}) = \sqrt{2S - a_{n}({\bf r})^{\dag}a_{n}({\bf r})} a_{n}({\bf r})$ with $a_{n}({\bf r})$ and $a_{n}^{\dag}({\bf r})$ for $n = 1,2,3,4$ denoting the four inequivalent sites of the unit cell, being the boson operators. 
Assuming $S \gg 1$, we obtain the Hamiltonian for non-interacting magnons written in Fourier space as
\begin{align}
\label{eq:matrix}
H &= J_{1}S \left[
\begin{array}{cccc}
 \lambda_{\bf k}+ \Delta_{\bf k} & {\tilde \gamma}_{\bf k} & t_{k_{z}} & 0 \\
{\tilde \gamma}_{-\bf k} & \lambda_{\bf k} - \Delta_{\bf k} & 0 & t_{k_{z}} \\
t^{*}_{k_{z}} & 0 &  -\lambda_{\bf k} + \Delta_{\bf k} & -{\tilde \gamma}_{\bf k} \\
0 & t^{*}_{k_{z}} & - {\tilde \gamma}_{-\bf k} &  -\lambda_{\bf k} - \Delta_{\bf k}  
\end{array}
\right]
\nonumber
\\
&
+\mu_{\mathrm{B}}BS+S (J_{\mathrm{a}}^{\prime} + J_{\mathrm{e}}^{\prime}),
\end{align}
where ${\tilde \gamma}_{\bf k}  =  2e^{i{\tilde k}_{x}\frac{1}{2\sqrt{3}}}\cos\left( \frac{{\tilde k}_{y}}{2}\right) + e^{-i{\tilde k}_{x}\frac{1}{\sqrt{3}}}$ with  ${\tilde k}_{x}  = k_{x} + \sqrt{3}\frac{D^{[\mathrm{R}]}}{J_{1}}m_{y}$ and ${\tilde k}_{y}  = k_{y} - \sqrt{3}\frac{D^{[\mathrm{R}]}}{J_{1}}m_{x}$. We also introduced 
$\Delta_{\bf k} = 2\Delta \left[\sin(k_{y}) - 2\sin\left( \frac{k_{y}}{2}\right)\cos\left(\frac{\sqrt{3}k_{x}}{2} \right) \right]$ with $\Delta =\frac{ D^{[\mathrm{z}]}}{J_{1}}m_{z}$, and $t_{k_{z}}= t_{\mathrm{a}}e^{ik_{z}} + t_{\mathrm{e}}e^{-ik_{z}}$ with $ t_{\mathrm{a}/\mathrm{e}} = \frac{J^{\prime}_{\mathrm{a}/\mathrm{e}}}{J_{1}}$. For the diagonal elements we introduce $\lambda_{\bf k} = \lambda - \zeta_{\bf k}$ with $\zeta_{\bf k} = 2\delta\left[ \cos(k_{y})+ 2\cos\left(\frac{\sqrt{3}k_{x}}{2} \right)\cos\left(\frac{k_{y}}{2} \right)  \right]$, where $\delta =\frac{ J_{2}}{J_{1}}$ and $\lambda = 3-6\delta$. 
The different signs in front of $\gamma_{\bf k}$ and $\lambda_{\bf k}$ for the top and bottom layers are due to the difference in the sign of the exchange interaction, i.e., ferromagnetic or antiferromagnetic. 
The space of the Hamiltonian is defined by the spinor $\Psi({\bf k}) = \left[ a_{1}({\bf k}),a_{2}({\bf k}),a_{3}({\bf k}),a_{4}({\bf k})  \right]^{\mathrm{T}}$.
A straightforward diagonalization of the Hamiltonian gives the energy spectrum
\begin{align}\label{spectrumweyl}
(\epsilon_{\pm})^2/(SJ_{1})^2  &=  
\lambda_{\bf k}^2 +  \Delta_{\bf k}^2 +  \vert t_{k_{z}} \vert^{2} + \vert\tilde{\gamma}_{\bf k}\vert^2 
\\
&
\pm 2 \sqrt{ \lambda_{\bf k}^2\left( \Delta_{\bf k}^2 + \vert \gamma_{\bf k} \vert^2\right) + \vert t_{k_{z}} \vert^{2} \Delta_{\bf k}^2}\,,
\nonumber
\end{align}
where we define magnon energy by $E_{\pm}$ with $\epsilon_{\pm} \equiv E_{\pm} - \mu_{\mathrm{B}}BS - S (J_{\mathrm{a}}^{\prime} + J_{\mathrm{e}}^{\prime})$.

In the following, we will search for the degeneracies (band touching) of the spectrum.
We note that they can occur only for the $\epsilon_{-}^2$ spectrum branch. 
It is straightforward to see that the degeneracy occurs at the ${\bf K}^\prime = (0,\frac{4\pi}{3})$ and ${\bf K}= (0,-\frac{4\pi}{3})$ points of the two-dimensional Brillouin zone, and at values of the $k_{z}$ determined from the following considerations.   
Close to the ${\bf K}^\prime$ point, we approximate ${\tilde \gamma}_{\bf k} \approx -\frac{\sqrt{3}}{2}({\tilde k}_{y} + i{\tilde k}_{x})$, $\Delta_{\bf k} \approx -3\sqrt{3}\Delta$, $\zeta_{\bf k} \approx -3\delta$, and $\lambda_{\bf k} \approx 3(1-\delta)$.
The points of possible degeneracy are defined by the equation
\begin{align}\label{condition}
\left( \Delta_{\bf k}^2 +\vert t_{k_{z}} \vert^2 + \vert {\tilde \gamma}_{\bf k}\vert^2 - \lambda_{\bf k}^2 \right)^2 
= 4\vert t_{k_{z}}\vert^2 \left( \Delta_{\bf k}^2 - \lambda_{\bf k}^2 \right).
\end{align}
 
Let us carefully analyze different cases of the parameters, in particular focusing on the strength of the DMI and inter-layer exchange interactions.

\begin{figure} 
\centerline{\includegraphics[clip, width=0.9  \columnwidth]{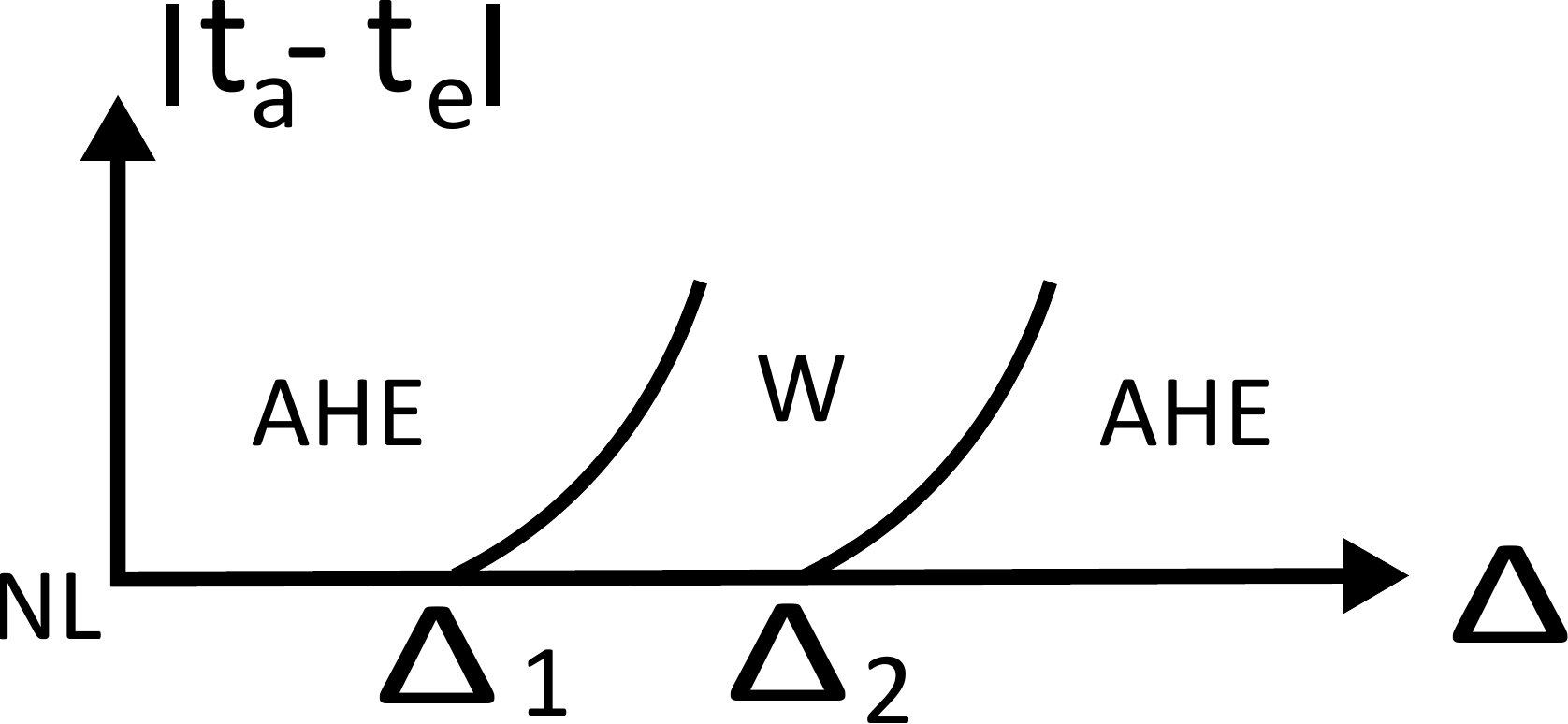}}

\protect\caption{Phase diagram of the model. The $X$-axis is the strength of the DMI, and the $Y$-axis is the difference of the inter-layer exchange interaction. Here: NL is the unstable nodal line, which occurs only when $t_{\mathrm{a}} = t_{\mathrm{e}}$ and $D^{[\mathrm{z}]} = 0$; AHE stands for the anomalous Hall effect magnons, and W stands for the Weyl magnons. Boundaries are defined by the conditions on the DMI strength derived in the text:
$\Delta_{1} = \frac{1}{\sqrt{3}}\vert 1-\delta\vert$ and $\Delta_{2} = \frac{1}{\sqrt{3}}\sqrt{(1-\delta)^2 + 4t_{\mathrm{e}}^2}$.  Schematics of the phase diagram are presented for $1-\delta = 0.5$ and $2t_{\mathrm{e}} = 0.5$.}

\label{figPhaseD}  

\end{figure}

(i) When $\Delta_{\bf k}^2 > \lambda_{\bf k}^2$, this inequality can be rewritten as $3\Delta^2>(1-\delta)^2$ at the ${\bf K}^\prime$ point (the same consideration applies to the ${\bf K}$ point), and Eq. (\ref{condition}) is reduced to $\vert {\tilde \gamma}_{\bf k} \vert^2 + \left( \sqrt{\Delta_{\bf k}^2 - \lambda_{\bf k}^2} - \vert t_{k_{z}}\vert \right)^2 = 0$. 
It is satisfied only when $\frac{1}{2\sqrt{ t_{\mathrm{a}}t_{\mathrm{e}} }} \sqrt{3\Delta^2 - (1-\delta)^2 - (t_{\mathrm{a}} - t_{\mathrm{e}} )^2 } < 1$, which sets in another condition for the DMI strength, namely $3\Delta^2 > (1-\delta)^2 + (t_{\mathrm{a}} - t_{\mathrm{e}} )^2 $. 
We derive the values of $k_{z}$ that nullify the bracket, and get $k_{z}^{\pm} = \pm \arccos\left( \frac{1}{2\sqrt{ t_{\mathrm{a}}t_{\mathrm{e}} }} \sqrt{3\Delta^2 - (1-\delta)^2 - (t_{\mathrm{a}} - t_{\mathrm{e}} )^2} \right)$. 
The condition ${\tilde \gamma_{\bf k}} = 0$ is easy to satisfy, and therefore we find two Weyl points at the ${\bf K}^\prime$ point, namely at $(0, \frac{4\pi}{3},k_{z}^{\pm})$ (for $D^{[\mathrm{R}]} = 0$).  See Fig.~(\ref{fig}) for the magnon spectrum in the simplified case.
When the condition  $\frac{1}{2\sqrt{ t_{\mathrm{a}}t_{\mathrm{e}} }} \sqrt{3\Delta^2 - (1-\delta)^2 - (t_{\mathrm{a}} - t_{\mathrm{e}} )^2 } > 1$ is satisfied the system is gapped, and the system is an analog of the AHE phase for magnons (magnon AHE phase).
This phase is discussed in Figs.~(\ref{figPhaseD}) and (\ref{fignew}).

(ii) When $\Delta_{\bf k}^2 < \lambda_{\bf k}^2$, the degeneracy occurs at $\vert t_{k_{z}}\vert = 0$ and at the points defined by the equation $\vert {\tilde \gamma_{\bf k}}\vert^2 =   \lambda_{\bf k}^2 - \Delta_{\bf k}^2$. 
The first condition is satisfied only when $t_{\mathrm{a}} = t_{\mathrm{e}}$. 
When both conditions are met, we obtain a nodal line touching of the spectrum close to the ${\bf K}^\prime$ point. 
The nodal line phase is unstable and it separates two distinct phases which are characterized by a surface state. 
Namely, when the DMI is absent, the surface Dirac state exists if the bulk is interrupted by breaking the largest of the two inter-layer exchange couplings, $J_{\mathrm{a}}^{\prime}$ and $J_{\mathrm{e}}^{\prime}$. 
This is consistent with the Shockley model discussed in Ref.~[\onlinecite{PhysRevB.86.075304}]. 
Since the same scenario occurs at both ${\bf K}$ and ${\bf K}^{\prime}$, we obtain two Dirac surface magnon states occurring at the ${\bf K}$ and ${\bf K}^{\prime}$ points.
Finite values of the DMI will gap the surface Dirac state, and one obtains the AHE magnons. 
Such transitions are shown in Fig.~(\ref{figPhaseD}).
We note that this behavior is expected as the above model is a honeycomb layer based magnon analog of a Weyl semimetal proposed in Ref.~[\onlinecite{BurkovBalentsPhysRevLett.107.127205}] and with details elaborated in Ref.~[\onlinecite{PhysRevB.85.165110}].

(iii) According to Fig.~\ref{fignew}, there is a special point $(0,0,\pi/2)$ in the Brillouin zone at which an accidental degeneracy occurs. 
Exactly at this point, the DMI vanishes $\Delta_{\bf k} =0$, a small $k$ expansion around the point gives $\Delta_{\bf k} \approx \frac{1}{4}\Delta k_{y}(3k_{x}^2 - k_{y}^2)$ and $\mathrm{Re}\tilde{\gamma}_{\bf k} \approx 3-\frac{1}{4}\tilde{k}^2$, and given $\Delta \ll 1$ we can neglect the DMI in the spectrum (see  Ref.~[\onlinecite{KovalevZyuzinLiPhysRevB.95.165106}] for details).  
The spectrum (for $D^{\mathrm{R}}=0$) is then $(\epsilon_{\pm})^2 = (\vert \lambda_{\bf k}\vert \pm \vert \gamma_{\bf k}\vert )^{2} +  \vert t_{k_{z}}\vert^2$, and there is a gap closing for $\epsilon_{-}$ at $k_{z}=\frac{\pi}{2}$ for the special case of $t_{\mathrm{a}}= t_{\mathrm{e}}$. 
According to Ref.~[\onlinecite{KovalevZyuzinLiPhysRevB.95.165106}], the Berry curvature at this point is defined by the DMI and is $\propto \Delta k_{y}^2 k_{x}^2$, which is not of a monopole type. Therefore, this point is not topological.

\section{Spin current due to the Berry curvature}
\label{section2}
In this section, we focus on the intrinsic transport properties of the Weyl and AHE magnons.
Intrinsic transport properties are those defined by the Berry curvature and are non-dissipative in nature. 
Therefore, we study the contributions from the points in the magnon Brillouin zone where the Berry curvature is the most singular, i.e., from the degeneracies. 
We simplify the model so the integrals can be calculated analytically. 
The prime task of the simplified model is to highlight the characteristic dependences of the intrinsic response. 
Also we would like to identify differences in the response structures of the Weyl and AHE magnons.
Numerical calculations for the full model are presented as well.

\subsection{Analytical results}
We consider a small-angle magnetization precession about the dc magnetic field that points in the $z-$ direction. 
A small magnetic field rotating about the $z-$ axis can be used to induce such precession.   
As shown in  Ref.~[\onlinecite{KovalevZyuzinLiPhysRevB.95.165106}], the dynamic $x-y$ part of the magnetization will cause spin and heat currents carried by the magnons to flow. 
In the following, we focus on the currents that are due to non-trivial topology of the magnon band structure. 
For the magnon particle current we obtain (see the Appendix for details)
\begin{align}\label{currentdynamics}
J^{[\mathrm{M}]}_{x} = \frac{1}{V} \frac{\sqrt{3} D^{[\mathrm{R}]}}{J_{1}} \sum_{\mu\nu}\sum_{\bf k} \Omega^{(\mu\nu)}_{xy}({\bf k})g(E_{\mu \nu} ) (\partial_{t} {\bf m})_{x},
\end{align}
where $\Omega^{(\mu\nu)}_{xy}({\bf k})$ is the Berry curvature, and $g(\epsilon) =(e^{\beta \epsilon} - 1)^{-1} $ is the Bose-Einstein distribution function with $\beta = J_{1}/T$.
One notices that a combination $\frac{\sqrt{3} D^{[\mathrm{R}]}}{J_{1}}$ is an effective charge of the magnons, while $(\partial_{t}{\bf m})_{x}$ is a fictitious electric field.
The opposite signs of $D^{[\mathrm{R}]}$ in the top and bottom layers of a unit cell, see Eq.~(\ref{ExchangeHamiltonian}), result in the same response of the magnons to the magnetization dynamics (the same sign would have resulted in a mutual compensation of the magnon response within a unit cell). 
Thus, the remaining part in Eq.~(\ref{currentdynamics}) has the meaning of the particle Hall conductivity of magnons,
\begin{align}\label{sigma}
\sigma_{xy} = \frac{1}{V}  \sum_{\mu\nu}\sum_{\bf k} \Omega^{(\mu\nu)}_{xy}({\bf k})g(E_{\mu \nu} ) .
\end{align}
Note that this response can also be associated with a spin Hall response defined by the spin Hall conductivity $\sigma_{xy}^s=-\hbar \sigma_{xy}$.
In order to make a comparison with the known anomalous Hall responses of Weyl semimetals in the case of fermions, 
we analytically estimate contributions from the Weyl points, where the Berry curvature is singular (of monopole type). 
We adopt a simplified model of Weyl magnons for which the spectrum is $\epsilon_{\mu \nu}=E_{\mu\nu} - h =  \mu v \sqrt{  k_{\parallel}^2 + ( \Delta_{\nu z} )^2}$, where $\mu = \pm$, $\nu = \pm$, $v \Delta_{\pm z} = \Delta \pm 2t\vert \cos(k_{z})\vert$ with $v=S$,  $\Delta =3\sqrt{3} D^{[\mathrm{z}]}/J_{1}$, $h = \mu_{\mathrm{B}}B/J_{1}$, and $t = J^{\prime}/J_{1}$. 
This is a spectrum of the model Eq. (\ref{spectrumweyl}) close to the ${\bf K}$ or ${\bf K}^{\prime}$ points in the hypothetical case with $\lambda_{\bf k} = 0$ and $t_{\mathrm{a}} = t_{\mathrm{e}} \equiv t = J^{\prime}/J_{1}$ (see Fig.~\ref{fig}). 
The Weyl points occur for the $\nu = -$ spectrum. 
For this model we derive the Berry curvature
\begin{align}
\Omega_{xy}^{(\pm ; \nu)}({\bf k}) = \mp \frac{1}{2} \frac{\Delta_{\nu z}}{( k_{\parallel}^2 + \Delta_{\nu z}^2  )^{3/2}},\label{Berry}
\end{align}
which is also shown in Fig.~\ref{fig}.
We calculate the current Eq. (\ref{currentdynamics}) in the limit $\beta h > 1$ and $\beta v \vert \Delta_{\pm z} \vert < 1$ (temperature larger than either the DMI strength or the inter-layer exchange interaction) to be
\begin{align}\label{resultcurrent1}
J^{[\mathrm{M}]}_{x} 
\approx 
 e^{- \frac{\mu_{\mathrm{B}}B}{T}} 
\frac{6\sqrt{3}D^{[\mathrm{z}]}}{\pi V T}   \ln\left[ \frac{\Lambda}{P} \right]
\left[  \frac{\sqrt{3} D^{[\mathrm{R}]}}{J_{1}}(\partial_{t} {\bf m})_{x}  \right].
\end{align}
where $P = \mathrm{max}(3\sqrt{3}D^{[\mathrm{z}]},2J^{\prime}) $ and $\Lambda < \mu_{\mathrm{B}}B$ is a cutoff (see the Appendix for details). Parameter $P$ distinguishes the two phases: the Weyl magnons when $3\sqrt{3}D^{[\mathrm{z}]}<2J^{\prime}$, and the AHE magnons in the opposite case.
We comment on the $\beta v \vert \Delta_{\pm z} \vert > 1$ case in the Appendix.

\begin{figure} \centerline{\includegraphics[clip, width=1  \columnwidth]{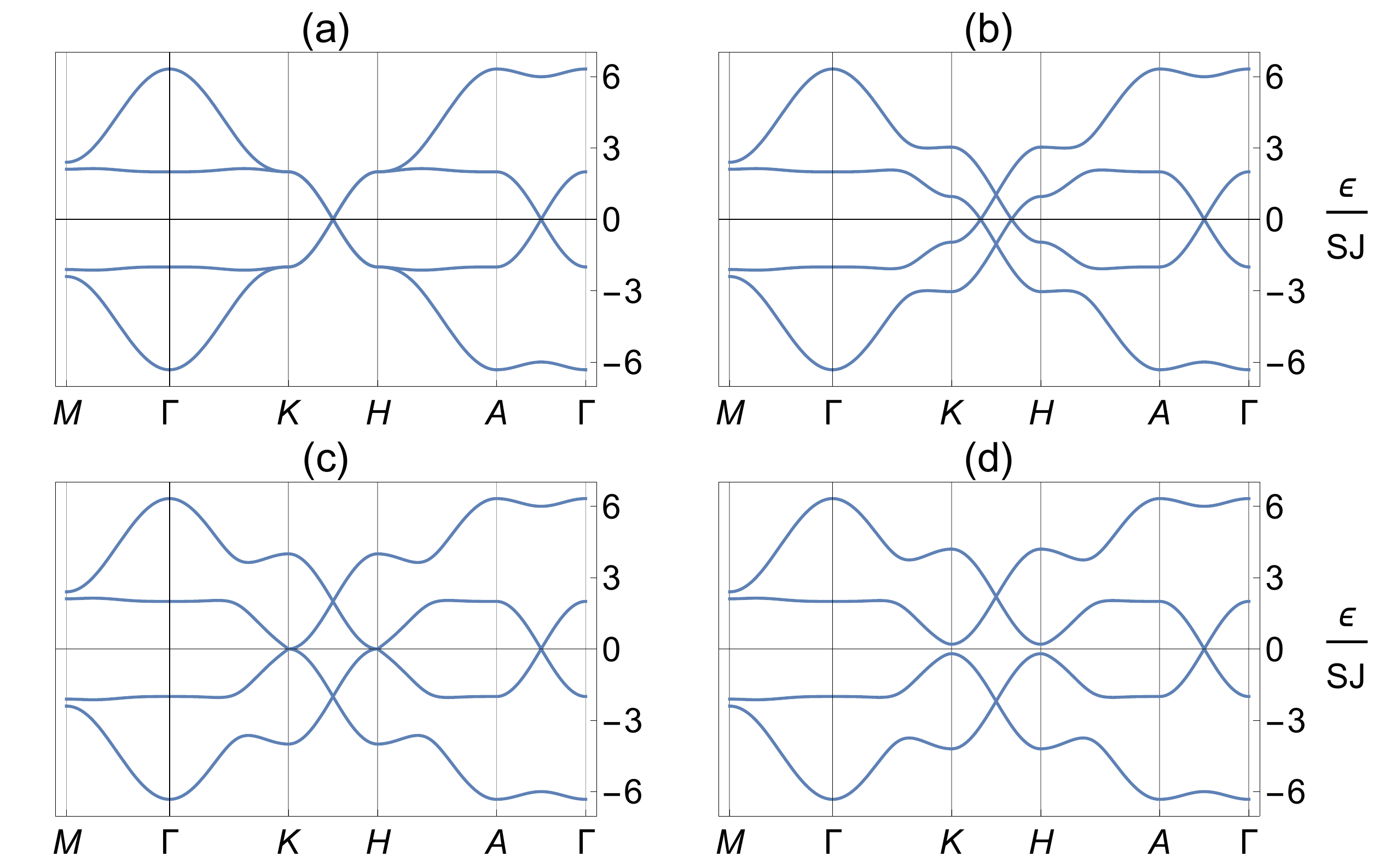}}

\protect\caption{(Color online) Spectrum of magnons $\epsilon(\boldsymbol k)$ in Eq.~(\ref{spectrumweyl}). (a) Vanishing DMI, $D^{[\mathrm{z}]}=0$, results in the formation of the Dirac node between the ${\bf K}$ and ${\bf H}$ points in the Brillouin zone. (b) DMI, $D^{[\mathrm{z}]}=0.2 J'$, splits the Dirac node into two Weyl points where the splitting is proportional to the strength of DMI. (c) and (d) The Weyl points annihilate at $3\sqrt{3}D^{[\mathrm{z}]}\geq 2J^{\prime}$ which leads to the formation of the AHE magnon phase. Here $J_2=J_1/3$, $J'=J_1$, and $J$ stands for $J_1$. We use the following notation for points in the Brillouin zone: $\Gamma=(0 ,0,0)$, $M=(\pi/\sqrt3 ,-\pi,0)$, $K=(0 ,-4\pi/3,0)$, $H=(0 ,-4\pi/3,\pi)$, and $A=(0 ,0,\pi)$. }

\label{fignew}  

\end{figure}

We now comment on the special point $(0,0,\frac{\pi}{2})$ (see Fig.~\ref{fignew}). The Berry curvature expanded close to this point in small $k$ is $\propto k^4$ (see Ref.~[\onlinecite{KovalevZyuzinLiPhysRevB.95.165106}] for details); therefore, the temperature behavior of the spin Hall response for small temperatures $T \ll J_{1}$ is $\propto \left(\frac{T}{J_{1}} \right)^7  e^{- \frac{\mu_{\mathrm{B}}B}{T}} $, and consequently, it is suppressed.

We can now see a difference between fermions and bosons.
Importantly, in the case of Weyl semimetals (fermions), the Hall conductivity is semi-quantized. 
This means that it is proportional to the splitting between the Weyl points in momentum space times the $\frac{e^2}{\hbar}$. 
In the case of bosons, as can be seen from Eq. (\ref{resultcurrent1}), the calculated anomalous response is also proportional to the splitting between the Weyl points, which is proportional to $D^{[\mathrm{z}]}$. However, the response is temperature dependent, $\propto\frac{1}{T}e^{- \frac{\mu_{\mathrm{B}}B}{T}} $, and hence by no means quantized. 
Because the Berry curvature for the $E_{\pm, \nu}$ energy bands is opposite in sign, the integrand defining the current in Eq.~(\ref{currentdynamics}) is less singular at the Weyl points for bosons compared to fermions. 
Technically, it is due to the vanishing of the difference of the Bose-Einstein distribution functions for $E_{\pm \nu}$ energy bands at the Weyl points, where $E_{- -}=E_{+ -}$. 
In the case of fermions, the contribution to the anomalous Hall response comes only from fully filled bands, say, $E_{- - }$ when Fermi energy is larger than $h$, and the cancellation of the Fermi distribution functions does not occur.

\begin{figure} \centerline{\includegraphics[clip, width=1  \columnwidth]{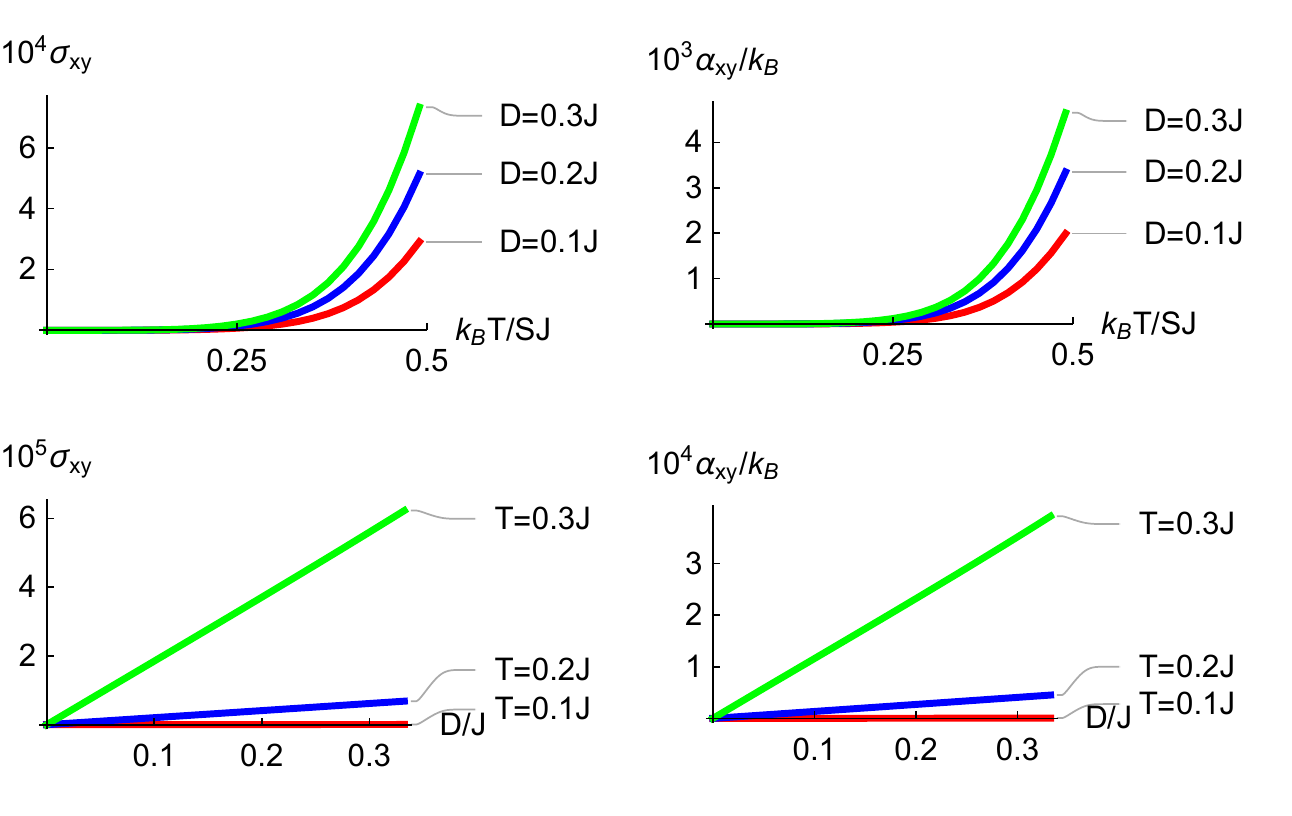}}

\protect\caption{(Color online) The Hall and spin Nernst responses of Weyl magnons. The top plots corresponds to values of the Dzyaloshinskii-Moriya interaction strength, $D^{[\mathrm{z}]}=0.1 J'$ (red), $D^{[\mathrm{z}]}=0.2 J'$ (blue), and $D^{[\mathrm{z}]}=0.3 J'$ (green). The bottom plots corresponds to values of temperatures, $T=0.1J$ (red), $T=0.2J$ (blue), and $T=0.3J$ (green). Here $J_2=J_1/3$, $J'=J_1$, $\mu_B B=13 J_1/2$, $D$ stands for $D^{[\mathrm{z}]}$, and $J$ stands for $J_1$.}

\label{fig2}  

\end{figure}

For the sake of generality, we also calculate the magnon spin Nernst current,
\begin{align}\label{alpha}
J_{x}^{[T]} = \alpha_{xy} ({\bm \nabla} T)_{y},
\end{align}
where $\alpha_{xy}$ is the spin Nernst coefficient and the temperature gradient is applied to the system in the $y$ direction, namely, $({\bm \nabla} T)_{y}$.
A generalization of the spin Nernst effect in fermion systems to boson systems, in particular to magnons in ferromagnets, was given in Ref. [\onlinecite{KovalevZyuzin2016}].  
The transverse to the temperature gradient response is again expressed via the Berry curvature, and it is given by 
\begin{align}\label{currentNernst}
\alpha_{xy} = \frac{1}{TV}  \sum_{\mu\nu}\sum_{\bf k} \Omega^{(\mu\nu)}_{xy}({\bf k})c_{1}\left(E_{\mu \nu} \right) ,
\end{align}
where $c_{1}(x) = \int_{0}^{x} d\eta~\eta \frac{dg(\eta)}{d\eta}$.
By studying the simplified model and after making the same approximations (see the Appendix for details), we obtain the expression for the spin density current,
\begin{align}\label{resultcurrent2}
J^{[\mathrm{T}]}_{x} \approx  &   \frac{\mu_{\mathrm{B}}B}{T} e^{-\frac{ \mu_{\mathrm{B}}B}{T}}
\frac{6\sqrt{3} D^{[\mathrm{z}]}}{\pi V T}   \ln\left[ \frac{\Lambda}{P}\right]
\frac{({\bm \nabla} T)_{y}}{T}.
\end{align}
Both currents, Eqs. (\ref{resultcurrent1}) and (\ref{resultcurrent2}), have similar features because of their dependence on the Berry curvature. The extra factor of $\frac{\mu_{\mathrm{B}}B}{T}$ in Eq. (\ref{resultcurrent2}) is due to the energy dependence of $c_{1}(x)$.

\subsection{Numerical results}
In Fig.~\ref{fignew}, we explore the magnon bands by varying the strength of DMI. We observe that the Weyl points split by DMI and the splitting in the momentum space is proportional to the strength of DMI. We further observe annihilation of Weyl points when condition the $3\sqrt{3}D^{[\mathrm{z}]}= 2J^{\prime}$ is satisfied. Increasing DMI further leads to the formation of the AHE magnons.
In Fig.~\ref{fig2}, we plot numerical results for the temperature and the DMI strength dependence of the magnon Hall and Nernst conductivities, corresponding to Eqs.~(\ref{sigma}) and (\ref{currentNernst}) for the model in Fig.~\ref{fig1}. Note that this calculation also accounts for the terms $\lambda_{\bf k}$ in Eq.~(\ref{eq:matrix}).  We observe that the effect rapidly increases with temperature which reflects the exponential dependence on temperature in Eqs.~(\ref{resultcurrent1}) and (\ref{resultcurrent2}). In addition, we observe a linear dependence on the DMI strength $D^{[\mathrm{z}]}$, again in agreement with Eqs.~(\ref{resultcurrent1}) and (\ref{resultcurrent2}). Nevertheless, we note that there can also be contributions due to the Berry curvature of other regions in the Brillouin zone, in addition to the contributions of the Weyl points discussed in the previous subsection. Note also that Fig.~\ref{fig2} corresponds to the Weyl magnon phase as in all plots $3\sqrt{3}D^{[\mathrm{z}]}<2J^{\prime}$.

\section{Conclusions}

To conclude, we have constructed a model of Weyl magnons, which is used to add new understanding of the structure and response functions of the Weyl magnons. 
Previous models utilized ferromagnets or antiferromagnets on pyrochlore, hyper-honeycomb, stacked honeycomb, and kagome lattices. \cite{WeylMagnons2016NatComm, PhysRevLett.117.157204, PhysRevB.95.224403,0256-307X-34-7-077501,Su.Wang:PRB2017,2017arXiv170802948J,OwerreArxiv2017a,OwerreArxiv2017b}

The honeycomb model in Ref. [\onlinecite{Su.Wang:PRB2017}] assumes anisotropic inter-layer exchange interaction which leads to the $\propto \sigma_{z}k_{z}$ term in the magnon Hamiltonian. 
This term is needed together with the dispersion linear in $k_{x}$ and $k_{y}$ space at the ${\bf K}$ and ${\bf K}^\prime$ points to fullfill the requirment of the Dirac Hamiltonian in three dimensions.
Our model is different from the one introduced in Ref. [\onlinecite{Su.Wang:PRB2017}], and it is based on stacked ferromagnet and antiferromagnet honeycomb layers. 
At zero external magnetic field, in general, there might not be a magnetic order due to frustration of the exchange interactions.  
The external magnetic field above the saturation level aligns all spins, and one can discuss magnons in such a system.
Different intra-layer exchange couplings (ferromagnetic/antiferromagnetic) create opposite magnon chiralities at the ${\bf K}$ or ${\bf K}^\prime$ points.
The interlayer exchange coupling then hybridizes opposite chiralities of the magnons at the ${\bf K}$ and ${\bf K}^\prime$ points.   
Furthermore, the DMI separates the opposite chiralities in momentum, thus creating the Weyl points. 
In the absence of DMI, the model can also realize surface states with the Dirac spectrum.
The model is a magnon analog of the fermion model given in Ref. [\onlinecite{BurkovBalentsPhysRevLett.107.127205}].

We have used the proposed model to calculate the intrinsic, due to the Berry curvature, responses of Weyl magnons. 
In particular, we have calculated responses to magnetization dynamics (magnon Hall effect) and temperature gradient (magnon Nernst effect) driven spin currents. 
The results are presented in the Eqs. (\ref{resultcurrent1}) and (\ref{resultcurrent2}). 
Using the similarity of our model to the fermion model given in Ref. [\onlinecite{BurkovBalentsPhysRevLett.107.127205}], we have compared the differences of the corresponding responses for magnons (bosons) and fermions (see the discussion after Eq. (\ref{resultcurrent1})).

\section{Acknowledgments}
We thank Bo Li for helpful discussions.
This work was supported by the DOE Early Career Award DE-SC0014189.

\appendix
\begin{widetext}
\section{Calculation of the spin current}
In the following, we will be using the notations in the main text.
We calculate a spin density current as a response first to the magnetization dynamics, and then to the temperature gradient. The latter is also called the magnon spin Nernst effect.

\subsection{Magnetization dynamics (magnon Hall effect)}
We assume the magnetic orderis in the $z-$ direction.
According to Ref.~[\onlinecite{KovalevZyuzinLiPhysRevB.95.165106}], the spin current flowing in the $x-$ direction driven by magnetization dynamics in the $x-y$ plane is 
\begin{align}
J^{[\mathrm{M}]}_{x} = \frac{1}{V} \frac{\sqrt{3} D^{[\mathrm{R}]}}{J_{1}} \sum_{\mu\nu}\sum_{\bf k} \Omega^{(\mu\nu)}_{xy}({\bf k})g(E_{\mu \nu} ) (\partial_{t} {\bf m})_{x},
\end{align}
where $\Omega^{(\mu\nu)}_{xy}({\bf k})$ is the Berry curvature, and $g(\epsilon) =(e^{\beta \epsilon} - 1)^{-1} $ is the Bose-Einstein distribution function with $\beta = J_{1}/T$.
We assume a simplified model for which the dimensionless spectrum is $E_{\mu \nu} = h + \mu v \sqrt{  k_{\parallel}^2 + ( \Delta_{\nu z} )^2}$, where $\mu = \pm$ denotes the upper/lower Dirac cones with respect to the energy parameter $h$, $\nu = \pm$ is the gapped/ungapped case, again, with respect to the energy parameter $h$, $v \Delta_{\pm z} = \Delta \pm 2t\vert \cos(k_{z})\vert$, and the dimensionless velocity is $v=S$.
Specifically, only the $\nu = -$ energy bands are degenerate at the Weyl points.
We assume that momenta are bound such that $h \gg v \sqrt{  k_{\parallel}^2 + ( \Delta_{\nu z} )^2}$. 
This is a spectrum close to the ${\bf K}$ or ${\bf K}^{\prime}$ points in the case when $\lambda_{\bf k} = 0$ and $t_{\mathrm{a}} = t_{\mathrm{e}} \equiv t = J^{\prime}/J_{1}$ for the model discussed in the text. 
We chose such parameters to highlight the differences in calculations of the anomalous Hall effect between known fermion Weyl systems and the present Weyl boson model. 
The identity
\begin{align}
\frac{1}{e^{\beta E_{+\nu}} -1} - \frac{1}{e^{\beta E_{-\nu}} -1} 
= \frac{\sinh(\beta v \sqrt{  k_{\parallel}^2 + \Delta_{\nu z}^2 })}{\cosh(\beta v \sqrt{  k_{\parallel}^2 + \Delta_{\nu z}^2 }) - \cosh(\beta h)}
\end{align}
is of use. 
We now calculate the current due to the Berry curvature in a case when analytic approximation is possible. 
The Berry curvature of the model for various bands is calculated as 
\begin{align}
\Omega_{xy}^{(\pm ; \nu)}({\bf k}) = \mp \frac{1}{2} \frac{\Delta_{\nu z}}{( k_{\parallel}^2 + \Delta_{\nu z}^2  )^{3/2}}.
\end{align}
We note that the Berry curvature is the same for both the ${\bf K}$ and ${\bf K}^\prime$ points.
The expression defining the current is
\begin{align}
&
2\frac{1}{2(2\pi)^2} \sum_{\nu}\int_{-\pi/2}^{\pi/2} \Delta_{\nu z} dk_{z} \int_{0}^{\Lambda} k_{\parallel} dk_{\parallel} \frac{1}{(k_{\parallel}^2 + \Delta_{\nu z}^2)^{3/2}} 
\frac{\sinh(\beta v \sqrt{  k_{\parallel}^2 + \Delta_{\nu z}^2 })}{\cosh(\beta v \sqrt{  k_{\parallel}^2 + \Delta_{\nu z}^2 }) - \cosh(\beta h)}
\\
&
= \frac{\beta v}{(2\pi)^2} \sum_{\nu}\int_{-\pi/2}^{\pi/2} \Delta_{\nu z} dk_{z} 
\int_{ \beta v \vert \Delta_{\nu z} \vert  }^{ \beta v\Lambda } 
\frac{dy}{y^2} \frac{\sinh(y)}{\cosh(y) - \cosh(\beta h)}
\\
&
\approx -\frac{2}{\cosh(\beta h)}\frac{\beta v}{(2\pi)^2} \sum_{\nu}\int_{-\pi/2}^{\pi/2} \Delta_{\nu z} dk_{z} 
\ln\left[ \frac{\Lambda}{ \vert \Delta_{\nu z}\vert} \right]
\\
&
\approx -2 e^{-\beta h}\frac{\beta \Delta}{\pi}   \ln\left[\frac{\Lambda}{\mathrm{max}(\Delta,2t)} \right],
\end{align}
where the factor of 2 in the first line is due to equal contributions from the ${\bf K}$ and ${\bf K}^\prime$ points.
Here the cutoff is $\Lambda > \mu_{\mathrm{B}}B$.
Going from the second line to third, we assumed that $\beta v \vert \Delta_{\nu z} \vert < 1$, and approximated the integral within logarithmic accuracy.  
Going from the third to fourth line, we again estimated the integral within logarithmic accuracy, and assumed $\beta h \gg 1$.
The current, recalling $\Delta = 3\sqrt{3} D^{[\mathrm{z}]}/J_{1}$, $h \approx \mu_{\mathrm{B}}B/J_{1}$, and $t \approx J^{\prime}/J_{1}$, is then
\begin{align}
J^{[\mathrm{M}]}_{x} \approx 
 e^{- \mu_{\mathrm{B}}B/T} 
\frac{6\sqrt{3}D^{[\mathrm{z}]}}{\pi V T}   
\ln\left[\frac{\Lambda}{  \mathrm{max}( 3\sqrt{3}D^{[\mathrm{z}]},2J^{\prime})} \right] 
\frac{\sqrt{3} D^{[\mathrm{R}]}}{J_{1}}(\partial_{t} {\bf m})_{x}.
\end{align}
We stress that the spin current of the magnetic system discussed in the main text, will contain contributions from all regions of the Brillouin zone.
In the expression above, we have considered only the contribution from the ${\bf K}$ and ${\bf K}^\prime$ points for a very special case of $\lambda_{\bf k} = 0$ and  $t_{\mathrm{a}} = t_{\mathrm{e}}$. 

In the $\Delta > 2t$ case, it is instructive to obtain a result for the spin current that is a sum of a number of stacked AHE Chern magnon layers.
This regime happens at small temperatures, $\beta v \vert \Delta_{\nu z} \vert > 1$:
\begin{align}
&
\frac{\beta v}{2(2\pi)^2} \sum_{\nu}\int_{-\pi/2}^{\pi/2} \Delta_{\nu z} dk_{z} 
\int_{ \beta v \vert \Delta_{\nu z} \vert  }^{ \beta v\Lambda } 
\frac{dy}{y^2} \frac{\sinh(y)}{\cosh(y) - \cosh(\beta h)}
\\
&
\approx
-\frac{1}{\cosh(\beta h)}
\frac{\beta v}{2(2\pi)^2} 
\sinh\left(\beta \frac{3\sqrt{3} D^{[\mathrm{z}]}}{J_{1}}\right)
\sum_{\nu}\int_{-\pi/2}^{\pi/2} \mathrm{sign}\left( \Delta_{\nu z}\right) dk_{z}
\\
& 
=
-\frac{1}{\cosh(\beta h)}
\frac{\beta v}{4\pi} 
\sinh\left(\beta \frac{3\sqrt{3} D^{[\mathrm{z}]}}{J_{1}}\right).
\end{align}
The response in this case is then
\begin{align}
J^{[\mathrm{M}]}_{x} \approx 2 e^{-\frac{\mu_{\mathrm{B}}B}{T}}\frac{SJ_{1}}{2\pi V T} 
\sinh\left(\frac{J_{1}}{T} \frac{3\sqrt{3} D^{[\mathrm{z}]}}{J_{1}}\right)
\frac{\sqrt{3} D^{[\mathrm{R}]}}{J_{1}}(\partial_{t} {\bf m})_{x}.
\end{align}

We stress that in all of the above calculations of the current we have focused primarily on the low $k$ contribution to the integrals.
This is the only contribution that distinguishes the Weyl and AHE magnons.

\subsection{Temperature gradient (magnon spin Nernst effect)}
Spin current due to the Berry curvature driven by the temperature gradient $\frac{(\nabla T)_{y}}{T}$ is (magnon spin Nernst effect)
\begin{align}
J^{[\mathrm{T}]}_{x} = \frac{1}{V}  \sum_{\mu\nu}\sum_{\bf k} \Omega^{(\mu\nu)}_{xy}({\bf k})c_{1}\left[g(E_{\mu \nu}) \right] 
\frac{(\nabla T)_{y}}{T}.
\end{align}
To extract the analytic results, we approximate  
\begin{align}
c_{1}\left[ g(\epsilon) \right] 
&= \left[ 1+g(\epsilon) \right] \ln\left[ 1+ g(\epsilon) \right] - g(\epsilon)\ln\left[ g(\epsilon) \right] 
\\
&=  \ln\left[ g(\epsilon) \right] + \beta\epsilon\left[ 1+ g(\epsilon)   \right] 
\\
&\approx e^{-\beta\epsilon}\left( 1+ \beta\epsilon  \right).
\end{align}
For our simplified model we approximate
\begin{align}
c_{1}\left[ g(E_{+\nu}) \right] - c_{1}\left[ g(E_{-\nu}) \right] \approx -2  \beta^2 h e^{-\beta h}  v\sqrt{  k_{\parallel}^2 + ( \Delta_{\nu z} )^2}.
\end{align}
We apply the same approximations as in the previous section, namely $\beta h \gg 1$ and  $\beta v \vert \Delta_{\nu z} \vert < 1$. 
We then get for the integral defining the spin Nernst current the expression
\begin{align}
&
2\frac{1}{2(2\pi)^2} \sum_{\nu}\int_{-\pi/2}^{\pi/2} m_{\nu z} dk_{z} \int_{0}^{\Lambda} k_{\parallel} dk_{\parallel} \frac{1}{(k_{\parallel}^2 + \Delta_{\nu z}^2)^{3/2}}
\left\{ c_{1}\left[ g(E_{+\nu}) \right] - c_{1}\left[ g(E_{-\nu}) \right] \right\} 
\\
&
\approx 2  e^{-\beta h}\frac{\beta^2 h \Delta}{2\pi}   \ln\left[\beta  \mathrm{max}(\Delta,2t) \right].
\end{align}
The spin Nernst current then reads
\begin{align}
J^{[\mathrm{T}]}_{x} \approx  \frac{\mu_{\mathrm{B}}B}{T} e^{-\frac{ \mu_{\mathrm{B}}B}{T}}
\frac{6\sqrt{3} D^{[\mathrm{z}]}}{\pi V T}   
\ln\left[\frac{\Lambda}{  \mathrm{max}(3\sqrt{3}D^{[\mathrm{z}]},2J^{\prime})} \right]
\frac{(\nabla T)_{y}}{T}.
\end{align}

In the $\Delta > 2t$ case, at small ($\beta v \vert \Delta_{\nu z} \vert < 1$) temperatures we get
\begin{align}
J^{[\mathrm{T}]}_{x} \approx 2\frac{\mu_{\mathrm{B}}B}{T} e^{-\frac{\mu_{\mathrm{B}}B}{T}}\frac{SJ_{1}}{2\pi V T} 
\sinh\left(\frac{J_{1}}{T} \frac{3\sqrt{3} D^{[\mathrm{z}]}}{J_{1}}\right)
\frac{(\nabla T)_{y}}{T}.
\end{align}

\end{widetext}

\def\urlprefix{}
\def\url#1{}

\bibliographystyle{apsrev}
\bibliography{MyBIB}

\end{document}